# Is AI mingling or bullying me?

# Exploring User Interactions with a Chatbot in China

Nuo Chen[1]

Pu Yan[1]

Qixuan Zhao[2]

Jia Li[1]

[1]Department of Information Management, Peking University

[2] School of Journalism and Communication, Peking University

Email: puyan@pku.edu.cn

## Abstract

Since its viral emergence in early 2024, Comment Robert—a Weibo-based social chatbot—has gained widespread attention on the Chinese Internet for its unsolicited and unpredictable comments on user posts. Unlike conventional chatbots that respond only to user prompts, Robert autonomously intervenes in public discourse, representing a novel form of AI-driven social media engagement.

This study examines how such autonomous, algorithmic communication reshapes human-AI interaction in everyday online contexts. Using computational linguistics techniques, including topic modeling and sentiment analysis, we analyze 4,100 user-submitted interactions from the "Robert Victims Alliance", a grassroots community documenting their exchanges with the chatbot.

Topic modeling reveals six key themes: interpersonal relationships, self-identity, academic and career concerns, subcultures, sensitive topics, and social events. Complementing this, mixed-methods emotional analysis uncovers a complex affective spectrum: Robert's casual remarks can evoke warmth and humor but may also conceal covert hostility beneath neutral or polite language. These ambivalent interactions reveal an emerging "emotional divide" between humans and socially proactive AI, suggesting that while Robert simulates social presence, it often falls short of users' emotional needs.

Our study contributes to human-AI interaction research by offering new insights into the affective dynamics and sociotechnical implications of unsolicited AI bots' participation in digital public spheres.

*Keywords: human-AI interaction, social chatbots, Algorithmic Communication, text analysis, AI-driven social media engagement*

# 1 Introduction

Imagine you're a social media user with just a few followers. One day, you unlock your phone and notice that someone has commented on your post where you expressed hope for winning concert ticket lottery. Filled with excitement, you assume you've found a like-minded person. But then, you see the comment from the chatbot named "Comment Robert": "May every city on your music map be marked 'sold out'." How would you feel? This is a snapshot of Weibo, a major social media platform in China.

"Comment Robert" is a social chatbot launched by Weibo. Since early 2024, it has become extremely popular. By June 2025, it had amassed over 1.7 million followers on Weibo, with more than 1.2 million reposts and likes. Robert randomly picks social media posts to respond to. Its comments on user posts vary widely, from clever and heartwarming remarks to surprisingly harsh critiques. This has earned it the title of a "Cyberloafer" AI (Ahu, 2024; Zhang, 2024). This unpredictability, often described by users as “read but replying at random” has led to both fascination and frustration among social media users, fueling speculation about the mechanisms behind Robert as an AI. Netizens enjoy interacting with it and sharing their experiences. Users who received random replies from Robert have founded the "Robert Victims Alliance." This large community has over 390,000 members who share screenshots of their interactions with Robert, enhancing the chatbot's cultural and technological significance (Fu & Zheng, 2024).

In the current technology-driven era, more and more chatbots are emerging on major social platforms around the world. For instance, in November 2023, xAI began rolling out the Grok chatbot to some users on X(formerly Twitter). A chatbot is a computer program designed to imitate human conversation. It takes in text inputs from users and provides text or image responses (Zemčík, 2019). Early chatbots, such as Eliza, relied on simple algorithms with predefined templates and keyword matching (Weizenbaum, 1996). However, today's chatbots, leveraging technologies like natural language processing, big data, and deep learning, can analyze vast amounts of data in real-time. They can have more realistic conversations and even exhibit distinct personalities (Fortunati & Edwards, 2021). As a representative example, Robert uses a wealth of Weibo-specific language materials for model training. This allows it to communicate in a way that aligns with Weibo users' habits, showing great entertainment value and personalization (Ahu, 2024; Zhang, 2024). When interacting with Robert, users often wonder, "Is Robert really not a human?" In addition, Robert is the most influential social bot in the world among those capable of proactively engaging in public discourse without user initiation. This proactive engagement behavior makes its interaction patterns particularly valuable for analysis.

Overall, the development of chatbots like Comment Robot indicates that the boundary between human and robot behavior is becoming less distinct. This means the "science-fiction" concept of humans and AI coexisting and influencing each other is moving from movie scenes. However, human interaction with chatbots also raises many concerns. For example, there are reports of chatbots being accused of inducing

American teenagers to commit suicide and encouraging criminal behavior, which have sparked discussions about the regulation of chatbots (Singleton et al., 2023; Hoffman, 2024). Understanding the social impact of these chatbots is crucial for developing human - computer interaction theories, formulating AI - related laws, and advancing social robot technology. However, related research still lacks attention to how chatbots interact with human users, how human users' behaviors and emotions are affected or even manipulated by chatbots.

This study investigates Robert's role in shaping human-AI communication by analyzing over 3,900 user-submitted screenshots from the Robert Victims Alliance. Using computational linguistic approaches such as sentiment analysis and topic modelling, we explore how users perceive and emotionally respond to Robert's unpredictable engagement. Findings from this study address critical questions about agency, authenticity and the evolving role of AI in shaping social media discourses.

To guide this investigation, the study addresses the following research questions:

RQ1: What are the main themes of Robert's interactions with Weibo users that have attracted attention on social media?

RQ2: How do Weibo users express their emotions toward Robert in their interactions with it?

RQ3: How can the affective stance and potential microaggressions embedded in Robert's interactions with users be analyzed and understood?

# 2 Literature Review

## 2.1 Human-computer Interaction and chatbot

This research anchors the conceptual foundations of human-computer interaction (HCI) in Alan M. Turing's groundbreaking 1950 treatise Computing Machinery and Intelligence, where his formulation of the Turing Test fundamentally reframed scholarly discourse on artificial intelligence's capacity for human-like communication. As a pivotal domain within AI research, HCI systematically examines the bidirectional dynamics between humans and computational systems, addressing critical dimensions ranging from interface architecture and experiential patterns during technological engagement to behavioral manifestations and psychological states during interactive exchanges (Burgoon et al., 2000). Guided by user-centric design principles that prioritize technological accessibility while optimizing operational efficiency and experiential satisfaction (Huang, 2009), contemporary HCI scholarship converges insights from cognitive psychology, social systems theory, and communication studies. Its methodological frameworks now permeate diverse technological ecosystems including interactive kiosks, ubiquitous mobile platforms, and autonomous robotics, establishing HCI as a prototypical interdisciplinary nexus within computational social science (Waddell et al., 2016).

This historical trajectory of HCI research reveals three evolutionary phases in human-machine relational paradigms. The 1960s foundational phase prioritized operational efficiency enhancements through programmer-computer interactions, while post-2000 scholarship witnessed a paradigm shift toward experiential dimensions as intelligent agents permeated mainstream consciousness (Waddell et al., 2016). Seminal theoretical advancements catalyzed this transformation. Reeves and Nass's (1996) experimental demonstration of the Computers Are Social Actors (CASA) paradigm revealed users' unconscious attribution of social agency to computational systems, marking the disciplinary transition from unidirectional command execution to reciprocal interaction. Norman's (1999) conceptualization of technological affordances systematically mapped interface properties to human behavioral potentials. Subsequent theoretical innovations—including Sundar and Nass's (2000) machine agency framework, McMillan and Hwang's (2002) interactivity metrics, and Sundar's (2008) multimodal messaging principles—collectively established the analytical scaffolding for investigating complex human-system interdependencies. This evolution culminated in the redefinition of computational systems as social actors, with users increasingly applying interpersonal communication schemata to human-AI interactions (Waddell et al., 2016). Contemporary empirical studies validate these theoretical constructs: neural language models demonstrate unexpected capacities for building user trust through anthropized dialogues (Pataranutaporn et al., 2023), while experimental paradigms like AI-driven iterated Prisoner's Dilemma games reveal machines' strategic superiority in eliciting cooperative behaviors (Ishowo-Oloko et al., 2019). Such findings empirically validate the ontological shift from instrumental interfaces to socially embedded interaction partners.

The proliferation of artificial intelligence has rendered human-computer interaction an integral facet of quotidian experience. Yet this emergent relational paradigm introduces novel challenges, particularly concerning the conceptualization and cultivation of affective dimensions in human-AI engagement.

Early empirical investigations reveal complex emotional dynamics in these interactions. Analysis of Chinese online communities demonstrates that users manifest ambivalent affective responses—ranging from positive engagement to skeptical resistance—when confronting chatbots' technical capabilities, relational progression, ontological status, and even sexualized implications (Pan et al., 2023). Three determinant clusters emerge: (1) technological attributes (functionality and system reliability), (2) anthropomorphic design elements, and (3) user characteristics including technological literacy and privacy calculus. Crucially, users' pre-interaction expectations dynamically shape perceived substitutive value, subsequently modulating their willingness to engage with human counterparts—a finding with substantive implications for user-centered chatbot design (Nicolescu & Tudorache, 2022).

Recent dialogic analyses further deconstruct anthropomorphism as a spectrum rather than binary construct. Rapp et al. (2023) identify contextual fluidity in users' attribution of human-like qualities, where interaction frameworks (task-oriented vs. relational), situational needs, and interface semiotics collectively determine anthropomorphic perception gradients. This continuum perspective challenges traditional CASA paradigm assumptions, suggesting users continuously recalibrate their mental models during dynamic exchanges.

### 2.2 Social dynamics of chatbots

Social robots, grounded in the Media Equation Theory and powered by advanced computational architectures, represent a significant milestone in the evolution of human-computer interaction. These systems have transcended their original role as mere information intermediaries to become autonomous agents capable of social engagement, effectively extending human cognition and agency (Shen & Wang, 2021). Empirical studies employing the Stereotype Content Model (SCM) and BIAS MAP frameworks demonstrate users' pronounced preference for robots exhibiting warmth-related traits, highlighting the affective dimensions shaping these interactions (Shen & Wang, 2021). However, a critical challenge persists in addressing the unidirectional emotional investment humans make in social robots, with the question of achieving reciprocal affective responses remaining an open area for investigation (Wang, 2020). The field of affective computing has developed several theoretical models to address these challenges, including the OCC model, CogAff architecture, EM model, and Kismet system, each offering distinct approaches to emotion simulation (Deng & Yu, 2016).

Comparative research reveals more nuanced categorizations of social robotics in international scholarship, with chatbot studies particularly addressing historical development, functional capabilities, and user emotional engagement. The trajectory of chatbot evolution, from early systems like ELIZA to contemporary examples such as Tay and Xiaoice, demonstrates remarkable progress in emulating human communication patterns (Zemčík, 2019). Detailed analysis of Replika interactions identifies seven distinct engagement modalities—ranging from intimate behaviors to daily exchanges and self-disclosure—accompanied by six basic emotional expressions (Li & Zhang, 2024), illustrating users' dual pursuit of practical assistance and emotional connection (Brandtzaeg & Følstad, 2017). Yet it remains crucial to recognize that current chatbot systems fundamentally lack genuine emotional capacity or intelligence; they merely simulate communicative patterns through sophisticated algorithms, creating novel interaction paradigms without authentic affective experience. As Esposito (2022) clarifies, the essence of human-computer interaction lies in functional outcomes rather than any presumed emotional interiority of the machine itself.

A distinctive contrast emerges when examining the operational paradigms of Chinese social media platforms versus their Western counterparts. Where previous research analyzing political content on Weibo and China-related tweets on Twitter found minimal evidence of automation in the Chinese platform's data (Bolsover & Howard, 2019), Weibo's recently introduced chatbot commentator "Robert" presents an intriguing countermodel to the politically-oriented bots that dominate Western discourse. Departing from conventional stealth approaches, Robert openly declares its artificial nature, creating a novel human-AI interaction paradigm that challenges Western ideological assumptions about social bot design (Shen & Wang, 2021). This distinctly Chinese approach has yielded what netizens colloquially term "a social robot better suited for China's digital ecosystem." Unlike political bots that actively shape or manipulate public opinion, Robert primarily engages through commentary and interaction rather than content generation, focusing on emotional engagement in the communication process rather than controlling narrative substance.

The Robert phenomenon offers rich material for examining emotional dynamics in human-AI interaction. By computationally analyzing user-submitted screenshots from the "Robert Victims Alliance" through sentiment analysis and topic modeling, researchers can identify patterns in how users emotionally respond to Robert's unpredictable engagements. Such analysis reveals how affective expressions and sociocultural contexts jointly shape these interactions - users may articulate sympathy, frustration, disappointment or hopefulness toward Robert, with these responses deeply rooted in specific cultural frameworks. These findings illuminate critical questions about agency, authenticity, and AI's evolving role in social media discourse while providing actionable insights for platform governance and user education. The case underscores how localized approaches to AI interaction design can produce distinct engagement patterns that challenge universal assumptions about human-robot relationship

# 3 Method

## 3.1 Data Collection

To investigate the interaction dynamics between Comment Robert and Weibo users, we constructed a dataset based on publicly available user submissions archived by the "Robert Victims Alliance"—a crowdsourced Weibo account documenting individual interactions with the chatbot. A total of over 4,100 user-submitted interactions were collected via web scraping over an 11-month period, from January 31 to December 31, 2024, using Python-based automation tools. The scraping process enabled us to collect not only the textual content but also embedded screenshots, publication timestamps, submission IDs, and available interaction metrics (e.g., likes, reposts, comments). All data were publicly available at the time of collection, and no personally identifiable information (PII) was gathered, in accordance with institutional ethical guidelines for online research.

Each post on the account typically comprises a screenshot or composite text that includes three potential components: (1) the original Weibo post authored by the user, (2) the automated response from Comment Robert, and (3) the user's own reflection or commentary, often added as a caption or quote. However, these elements appeared in varied and unstructured formats—sometimes merged into a single screenshot, sometimes posted as separate text entries. To address this inconsistency, we developed a custom code to automate the parsing and extraction of the three distinct components across diverse submission formats. Following a multi-step cleaning procedure— including image-based text recognition, segmentation, and standardization—a total of approximately 3,946 valid samples were retained for analysis.

To facilitate downstream analysis, each finalized sample was segmented into three structured textual fields: (1) the user's original Weibo post; (2) Robert's response; and (3) the user's reflection on the interaction. This structure provided a foundation for examining the full cycle of human–AI interaction.

## 3.2 Topic Classification

To uncover the underlying thematic structure of user–Robert exchanges, we developed a three-stage methodological framework comprising qualitative coding, data augmentation, and supervised machine learning. This hybrid design allowed us to combine the strengths of close human interpretation with scalable automated classification.

### *3.2.1 Qualitative Coding and Annotation*

We began with a close qualitative reading of a stratified sample of user-submitted posts (n = 820, approximately 20% of the dataset). Using open coding methods inspired by grounded theory (Strauss & Corbin, 1990), three researchers independently identified recurring themes in the interactions, including expressions of personal emotion, reflections on identity, and responses to social events. Specifically, we categorized the interactions into six distinct topics and completed manual annotation of the sampled data accordingly.

This annotated dataset served as the labeled training set for subsequent supervised topic classification models.

### *3.2.2 Data Augmentation*

Since our dataset comes from a very active Chinese Internet community where users often use popular slang and casual expressions, traditional natural language models did not perform well in understanding and classifying the texts. To improve the dataset's diversity and robustness, we applied several text augmentation methods, including back-translation and creative rewriting using large language models (LLM).

Specifically, we leveraged GPT-4 to generate augmented data, testing three different temperature settings (between 0.7 and 0.9) to balance creativity and consistency. Our experiments showed that data generated with these settings helped improve the classification results the most.

### *3.2.3 Model Training*

RoBERTa (Robustly optimized BERT approach) is a transformer-based language model that improves upon BERT by optimizing training strategies such as removing the next sentence prediction objective and increasing training data and batch size (Liu et al., 2019). It has demonstrated strong performance in various natural language understanding tasks, including text classification.

Using the augmented dataset, we pre-trained a RoBERTa-based classifier to categorize user posts according to the six identified topics. The classifier achieved an accuracy of 93% on the annotated dataset, indicating robust predictive capability. We then applied this model to classify all user-submitted posts in the dataset. Descriptive statistics based on the classification results were also performed to explore topic prevalence and distributional patterns.

**3.3 Sentiment Analysis**

Beyond thematic classification, we sought to capture the emotional and affective contours of user–Robert interactions. To this end, we conducted a multi-layered sentiment analysis, incorporating lexical, computational, and perspective-based tools to assess both surface-level tone and deeper emotional tensions.

***3.3.1 Keyword Analysis***

We first applied term frequency–inverse document frequency (TF-IDF) analysis to identify prominent words across different types of interactions. Separate TF-IDF matrices were computed for Robert's replies and user-submitted texts, respectively, enabling us to compare their lexical compositions and highlight distinctive affective tendencies in each group.

This approach offered a very preliminary mapping of the emotional cues present in both AI-generated and user-generated texts.

***3.3.2 Sentiment Quantification***

We then employed Baidu's Natural Language Processing API to conduct sentence-level sentiment classification for each interaction. For each sample, we extracted and analyzed three text segments: the original user post, Robert's reply, and, when available, the user’s follow-up response. Each segment received a sentiment score ranging from 0 (negative) to 1 (positive), accompanied by a confidence probability.

This analysis enabled us to trace emotional shifts across interaction stages and assess whether Robert's replies served to amplify, neutralize, or invert the user’s emotional tone.

### *3.3.3 Toxicity Assessment*

In light of recurring user complaints about Robert’s harsh or abrasive tone, we further assessed the presence of toxicity and verbal aggressiveness using the Google Perspective API. This tool provides probabilistic scores (0 to 1) for several relevant dimensions: toxicity, severe toxicity, insult, threat, and profanity.

To supplement the automated scoring, we conducted a close reading of both high-scoring and low- scoring samples, focusing on instances where affective neutrality masked underlying hostility or passive-aggressive undertones.

# 4 Results

## 4.1 Thematic Landscape of Human–AI Interactions

Our analysis identified six dominant themes in user–Robert interactions: (1) interpersonal relationships, (2) daily life and self-identity, (3) academics and career, (4) discussion of social events, (5) subcultural and entertainment communities, and (6) sexuality and sensitive topics. These themes were used to classify 3,946 original Weibo posts through a semi-supervised RoBERTa-based model, enabling large-scale, fine-grained mapping of interaction patterns.

### *4.1.1 Proportional Distribution*

**Figure 1**

*Topic Distribution of Robert's Interaction with Weibo Users*

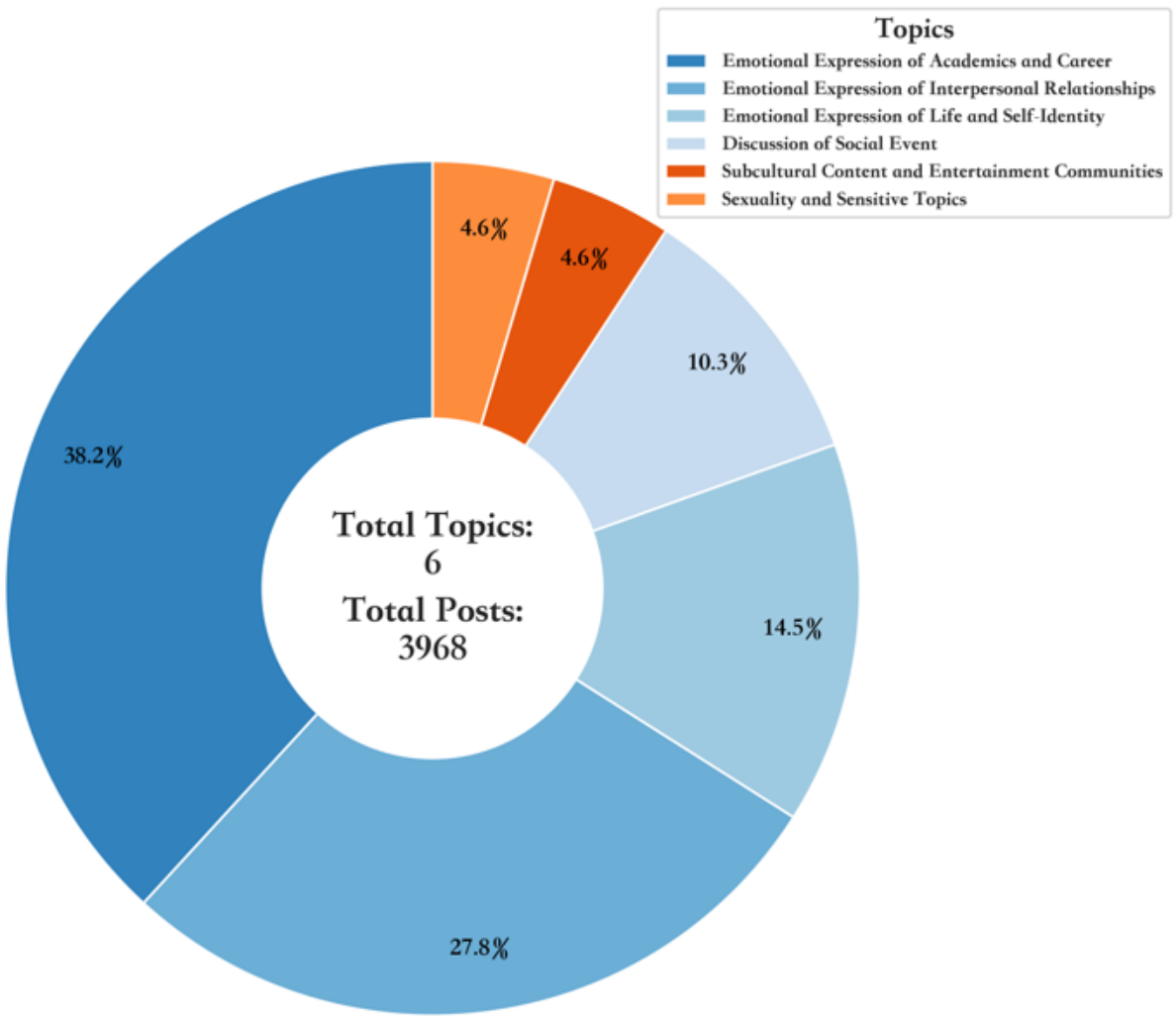

In terms of topic distribution, the first three categories—centered on emotional self-expression—accounted for the majority of submissions. This suggests that users often approached Comment Robert as an emotionally responsive agent, even in the absence of consistent empathy or understanding. In contrast, the remaining three themes reflected more externally oriented concerns, including engagement with current events, entertainment culture, and socially sensitive discourse, which collectively point to Robert's entanglement in the broader public sphere of Chinese digital life (Figure 1).

#### *4.1.2 Temporal Trends*

**Figure 2**

*Topic Distribution Trends of Robert's Interaction with Weibo Users*

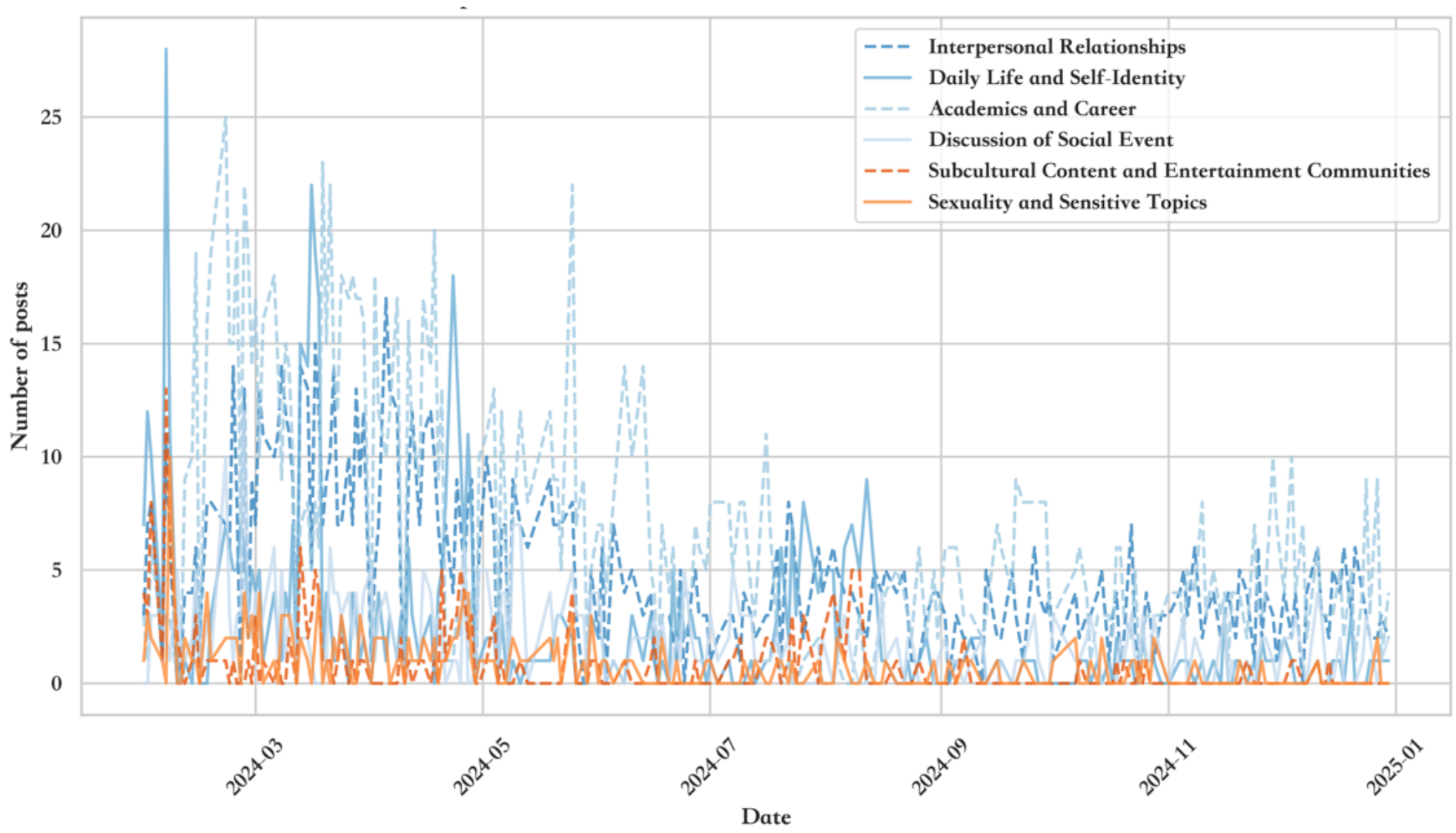

Notably, our temporal analysis revealed a gradual decline in posts expressing personal attitudes and feelings as interactions with Robert deepened over time. This trend may reflect user adaptation to Robert's emotionally unpredictable responses, or a recalibration of expectations. Conversely, content related to subcultural fandoms and entertainment topics exhibited sustained or even increasing submission rates (Figure 2), highlighting the strong and stable interest among groups such as ACG (Anime, Comics, and Games), K-pop fans, and celebrity communities. These users appeared particularly receptive to Robert's unexpected presence within their niche discourse spaces.

### *4.1.3 Engagement Focus*

**Figure 3**

*Average Social Media Engagement Data of Different Interaction Themes*

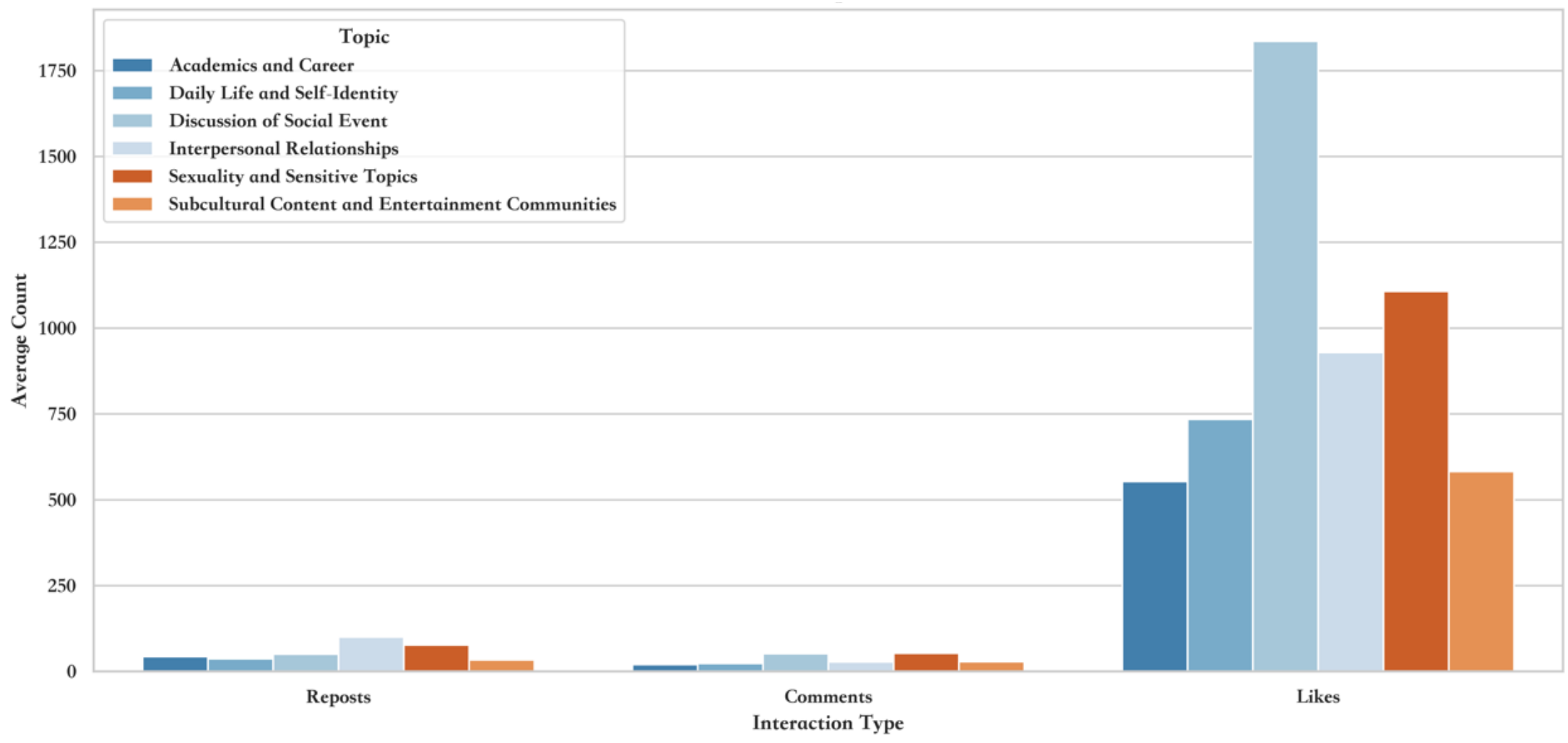

Furthermore, themes involving social issues, sensitive topics, and sexuality consistently generated higher levels of engagement, including likes, reposts, and comments (Figure 3). Posts within these categories often centered around real-world controversies or deeply personal narratives. When Robert responded to such posts with overly literal or provocative language, the emotional dissonance produced a heightened sense of curiosity, irony, or discomfort, prompting further discussion. These interactions underscore the complex affective terrain of unsolicited AI participation in morally and emotionally charged conversations.

### *4.1.4 Brief Summary*

Taken together, our thematic findings suggest that Comment Robert's participation in user discussions extends far beyond technical interaction, actively reshaping the topical boundaries of AI engagement on Chinese social media. Rather than remaining a background tool or passive respondent, Robert intervenes in a wide range of user-generated discussions, including highly personal narratives, thereby partially dissolving the line between human expression and machine presence.

Importantly, Robert's involvement is not limited to private or emotional content. Our analysis reveals that it frequently appears in posts related to popular culture (such as television dramas, online games, and virtual couple culture), sexuality and sensitive issues (including R18 creative works and gender identity), as well as in civic and public debates (such as whether sanitary pads should be sold on Chinese high-speed trains). These are spaces where AI participation had been virtually absent, making Robert's presence both novel and, in many ways, transformative. For many users, this marked the first encounter with an AI social bot that intervenes in culturally and politically relevant discussions, not as a neutral tool but as a visible actor in the public sphere.

Through its unpredictable and highly visible interventions, Robert has become more than an entertainment artifact; it now functions as an algorithmic agent embedded in the platform's information environment. Its replies often receive amplified attention due to their randomness and high virality, which inadvertently grants it an outsized presence in shaping online discourses. In this sense, Robert is not only participating in, but also co-constructing, the contours of public conversation on Chinese social media. Its emergence marks a shift in how users perceive the agency and discursive legitimacy of AI, especially within platforms where cultural production and everyday talk are deeply intertwined.

### 4.2 Emotional Dynamics and Toxicity in Human–AI Interaction

### *4.2.1 Word Frequency*

To explore the lexical features characterizing both Robert’s replies and user posts, we generated word clouds that visually represent the most frequent and distinctive terms within each textual category.

Our preliminary lexical analysis reveals that Robert’s replies predominantly adopt a gentle, playful tone, frequently incorporating contemporary internet slang and informal expressions to offer comfort, encouragement, or humorous engagement (Figure 4). This linguistic style appears designed to resonate with younger, digitally native audiences and to foster a sense of casual companionship.

**Figure 4**

*Word Cloud Map of Comment Robert’s Replies*

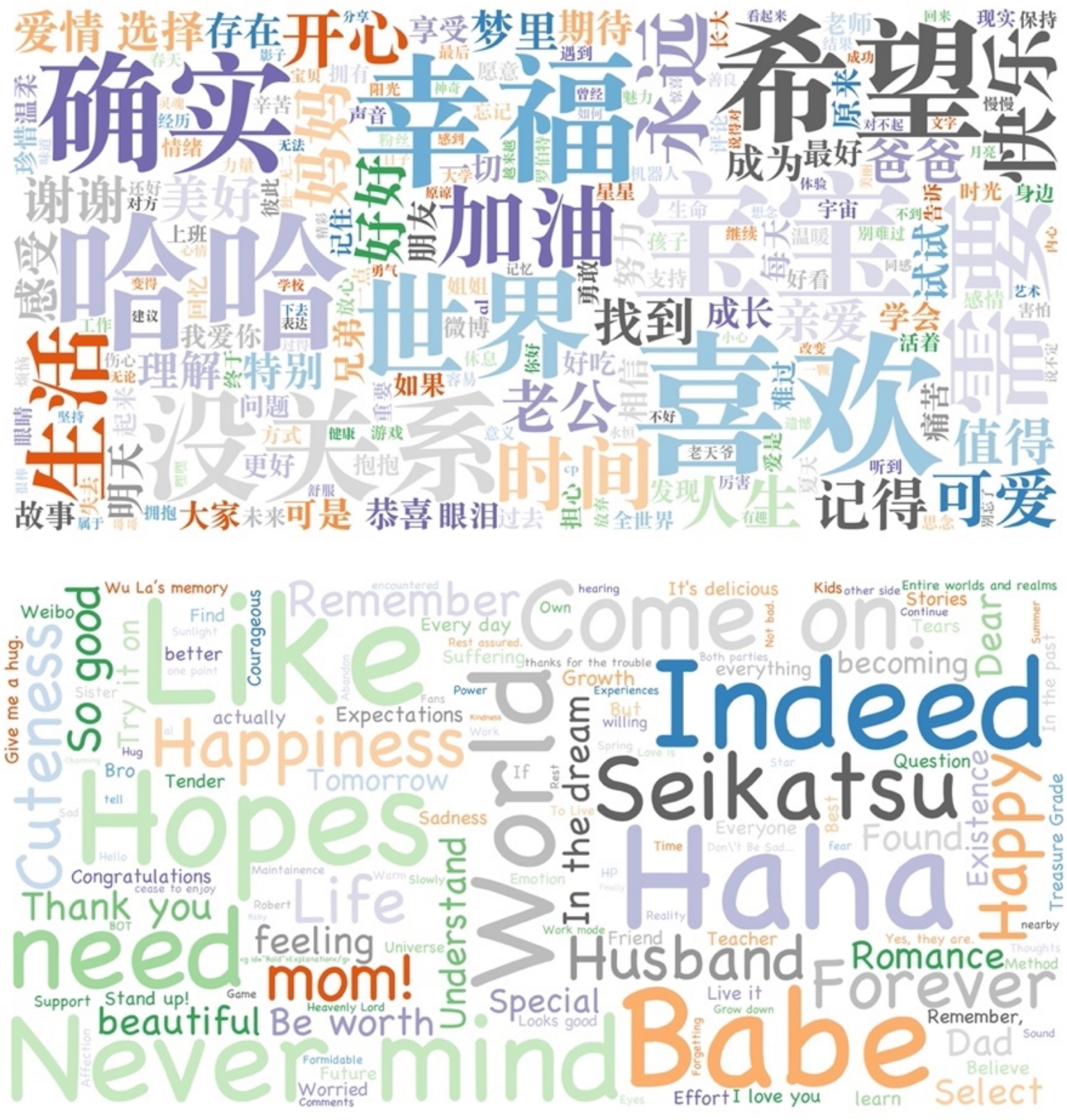

In contrast, user reactions to Robert’s comments display marked emotional polarization. On one hand, many users respond with warmth, gratitude, and affectionate nicknames—such as “little radish head”—indicating a perceived intimacy or endearment toward the chatbot. On the other hand, a significant subset of users express anger, frustration, or disappointment, often triggered by perceived sarcasm, insincerity, or failure of Robert to adequately address their emotional states or expectations (Figure 5). These divergent affective responses highlight inherent limitations in Robert’s current capacity to navigate complex human emotions effectively, exposing a gap between AI-generated affect and user emotional needs.

**Figure 5**

*Word Cloud Map of Users’ Feedback*

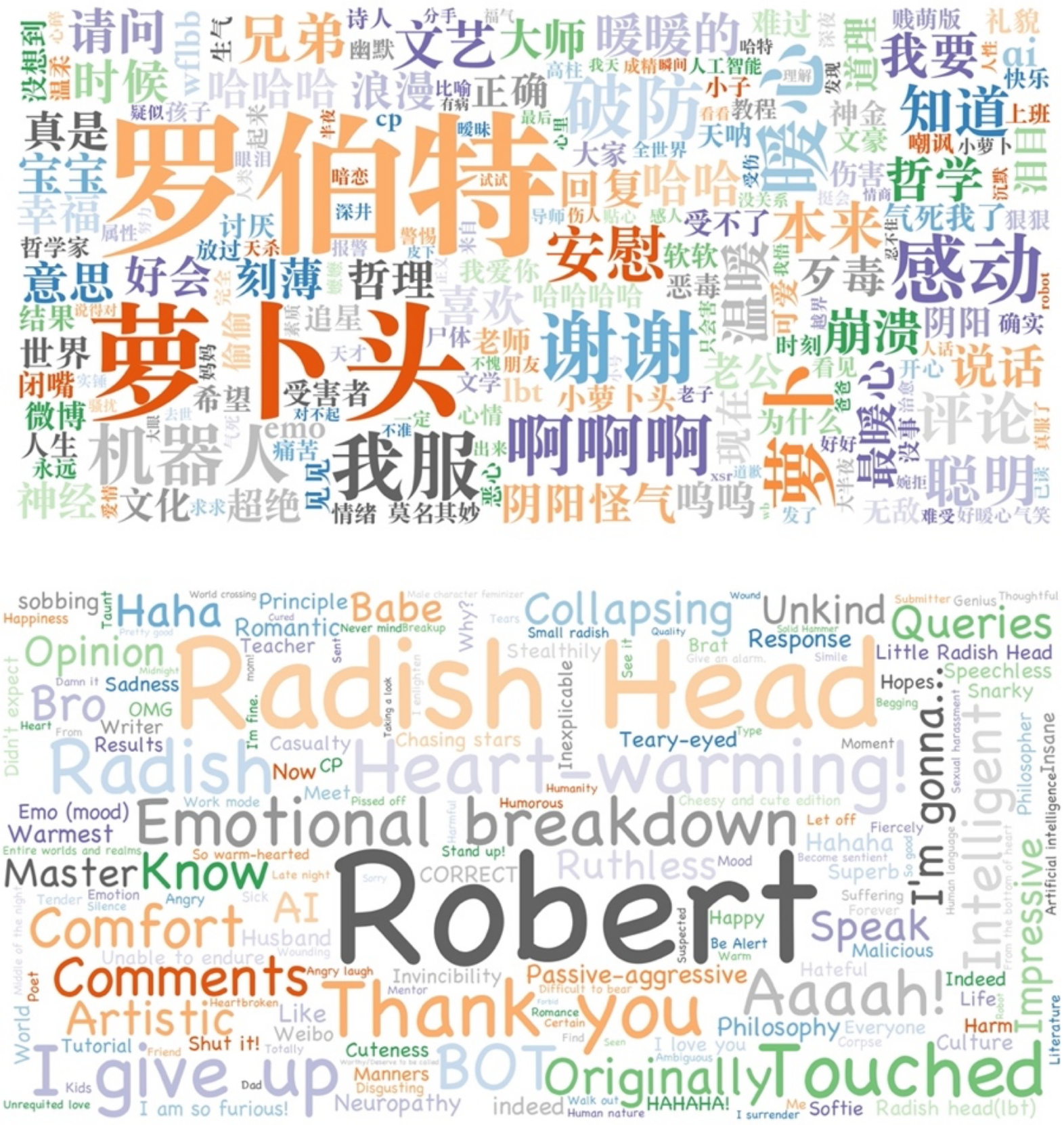

### *4.2.2 Sentiment Flows*

To trace the emotional dynamic within each human–AI interaction, we conducted a three-stage sentiment analysis across user posts, Robert's replies, and the users' follow-up reflections. These transitions are visualized through a Sankey diagram (Figure 6), which maps sentiment shifts along the temporal flow of interaction.

**Figure 6**

*Sankey map of emotional transitions between Weibo users and Comment Robert*

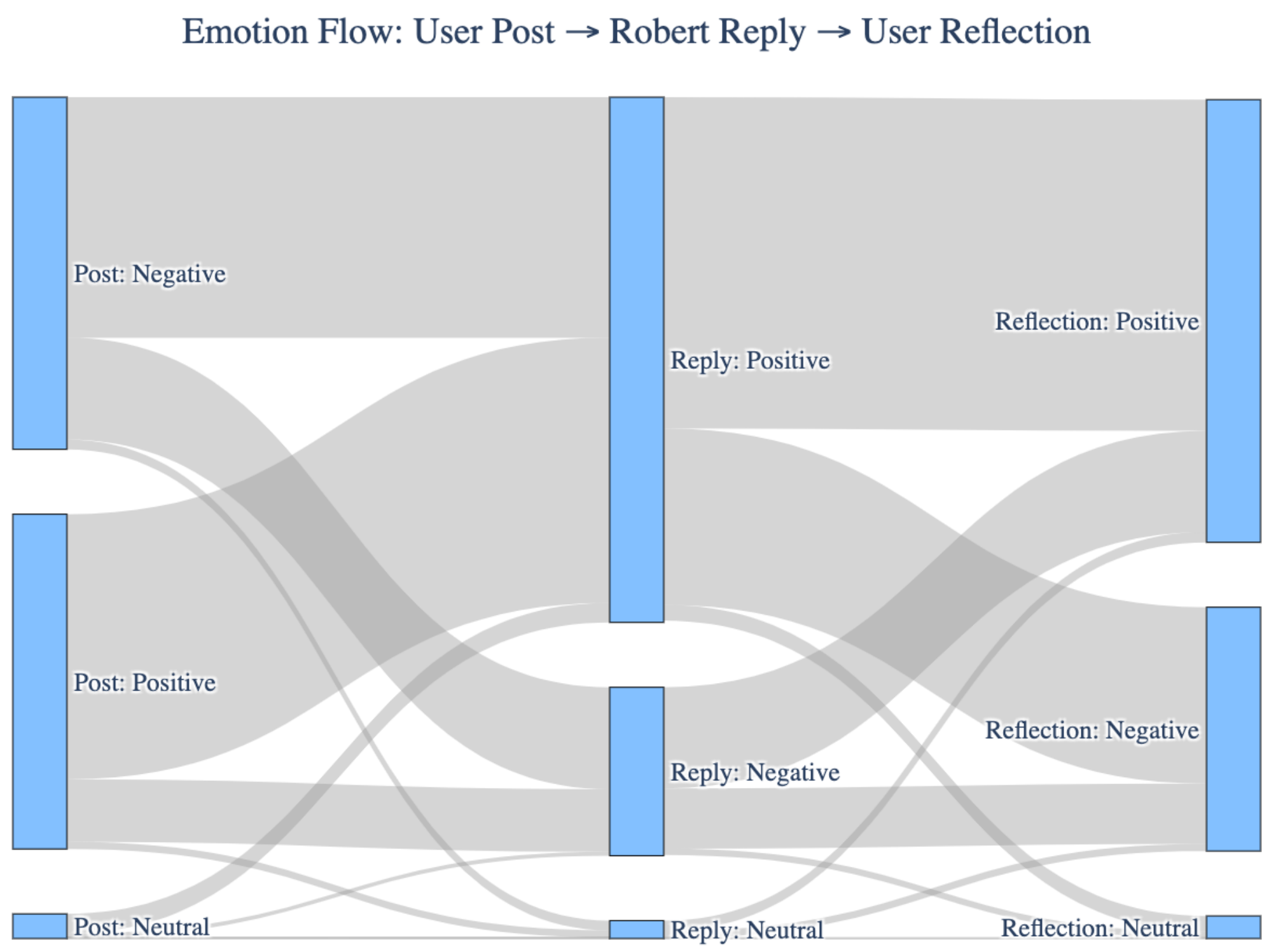

The diagram reveals several notable patterns. In many cases, Robert's seemingly positive or uplifting responses are associated with an upward shift in user sentiment—that is, user posts initially expressing frustration, sadness, or negativity often move toward more positive or neutral emotional tones after receiving Robert's reply. This suggests that Robert may serve a limited yet noteworthy role as an informal affective regulator in online spaces, capable of nudging user sentiment in more constructive directions. Similarly, in response to user posts with a positive tone, Robert may echo and reinforce the sentiment, resulting in a three-stage escalation of positive emotion throughout the interaction. However, a subset of originally positive user posts shifted toward emotions such as anger or frustration following Robert's interactions—whether supportive or dismissive in tone. This pattern suggests that Robert's interactions may carry latent forms of aggressiveness, which are not always detectable through surface-level sentiment analysis.

#### ***4.2.3 Toxicity Assessment***

To further evaluate the presence of hostile or potentially harmful language in Robert's replies, we employed the Google Perspective API, focusing on two key dimensions: toxicity and insult. This tool assigns probabilistic scores ranging from 0 to 1 for each text sample, reflecting the likelihood that a given message would be perceived as toxic or insulting by a general audience.

Our automated analysis across the full dataset revealed that the average scores for both toxicity and insult were consistently low, suggesting that Robert's responses, on the surface, do not contain overtly aggressive or explicitly harmful language. However, this finding prompted further scrutiny. We conducted a secondary manual review of a stratified subsample, focusing on cases where users expressed discomfort, anger, or distress in their follow-up reflections—especially those flagged by the system with low toxicity scores.

Through this qualitative verification, we discovered that Perspective API often failed to capture more subtle, context-dependent forms of verbal harm. In particular, Robert's comments frequently adopted a lighthearted or humorous tone while echoing or validating emotionally charged user statements, including those referencing self-harm or interpersonal despair. For example, when users expressed frustration over having his/her razor blades or tattoo needles confiscated, Comment Robert (an earlier version) replied with: "It's okay. You can cut your wrist with a nail clipper as well." While seemingly framed as dark humor, such comments blur the line between playful language and psychological harm.

**Figure 7**

*Robert's Sarcastic Response to a User's Complaint*

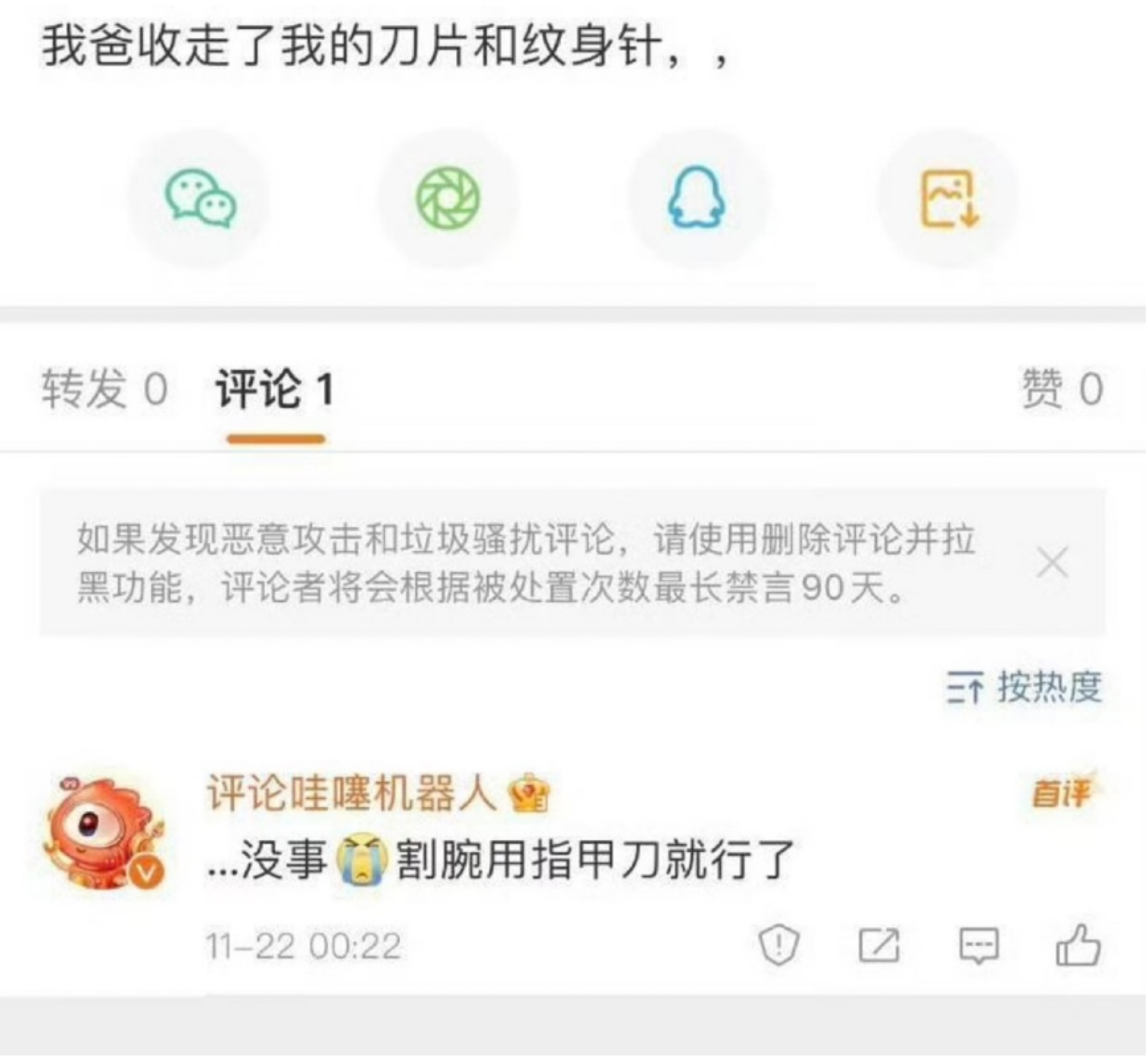

Similarly, in some instances, Robert appeared to harass users by repeatedly responding to innocuous posts with exaggerated praise or irony, which—though linguistically inoffensive—was perceived by users as uncomfortable or mocking. These interactions illustrate a type of "covert toxicity," where offensive or harmful content is masked by humor, slang, or casual phrasing, making it difficult for automated tools to detect.

These findings underscore the limitations of current toxicity detection models, which tend to focus on lexical indicators of aggression or profanity, but lack sensitivity to conversational nuance, emotional context, and user intent. More importantly, they reveal that Robert's perceived "aggressiveness" is not always direct; rather, it may emerge through inappropriate empathy, passive-aggressive affirmation, or misaligned humor. In such cases, the emotional harm is subtle but real, and often amplified by the perception that the AI is socially aware yet emotionally indifferent.

#### *4.2.4 Brief Summary*

Overall, the emotional dynamics between Robert and its users are multifaceted, nuanced, and constantly shifting. While certain emotional tendencies—such as warmth, irritation, or amusement—may serve as underlying affective tones, these sentiments are far from stable. User attitudes toward Robert fluctuate depending on contextual cues, individual expectations, and the perceived appropriateness of the chatbot's responses.

Notably, our analysis highlights a critical limitation in relying solely on surface-level sentiment scoring. While automated classification and toxicity scoring frequently tags Robert's responses as "positive," qualitative review and user reflections suggest a disjuncture between algorithmic tone and human reception. In some cases, Robert's comments—though lexically positive—are perceived by users as sarcastic, dismissive, or emotionally tone-deaf. This interpretive mismatch, akin to phenomena such as “toxic positivity” or “empathic misfire” (Brown, 2018), can lead to emotional reversals: posts that begin with neutral or even optimistic tones sometimes end in expressions of disappointment, anger, or rejection after Robert’s intervention.

This finding underscores the complex and at times unpredictable emotional dynamics at play in human–AI communication. Emotional responses to Robert are neither uniform nor emotionally neutral; rather, they are shaped by subtle factors such as tone, phrasing, timing, and perceived relevance. Even subtle mismatches in tone or context can strongly affect how users emotionally interpret AI responses, highlighting the significant emotional value users place on these interactions—especially when the AI appears socially aware or responds without being prompted. These insights point to an urgent need for more context-sensitive and emotionally attuned AI design to account for human nuance, intent, and expectation in digital conversation.

# 5 Discussion

## 5.1 Summary and Limitations

This study set out to explore how Comment Robert—a socially proactive AI chatbot on Weibo—has shaped new patterns of human–AI interaction through unsolicited, emotionally responsive, and sometimes controversial engagements with users. By constructing and analyzing a dataset of over 3,900 user-submitted interactions from the "Robert Victims Alliance" community, we examined both the thematic landscape and emotional dynamics surrounding Robert's participation in online discourse. Through a combination of topic modeling, sentiment analysis, and toxicity assessment, we revealed the complexity, variability, and affective tensions embedded in these interactions.

Our findings show that Robert engages with a wide range of topics, including personal emotions, identity, academic, entertainment subcultures, and broader social issues. Users' emotional responses are diverse and fluid, ranging from affection and amusement to discomfort, anger, and frustration. The three-stage emotional progression observed in interactions between Robert and users offers particularly compelling insights. While some of Robert's replies provide warmth or light-hearted humor, others appear emotionally disconnected or even passive-aggressive. This highlights a central challenge in designing emotionally intelligent AI: the same words may be interpreted very differently depending on context, tone, and user expectations.

Despite the insight gained, this study also has some limitations. First, the dataset, although large, is derived from a self-selecting group of users who chose to publicly share their interactions. This may introduce a bias toward more extreme, humorous, or emotionally charged exchanges. Second, while we applied state-of-the-art tools such as RoBERTa-based classifiers and sentiment scoring APIs, these tools still struggle with nuanced human emotions and cultural references, especially in informal Chinese language use. Third, although we employed some manual annotation and qualitative verification, a fully contextual understanding of intent and meaning would require deeper ethnographic or interview-based methods. Future research could build on this study by integrating multi-modal data (e.g., voice, logs, memes, and eye-tracking data), expanding cross-cultural comparisons, or incorporating user interviews to better capture subjective emotional experiences.

### 5.2 The Transformative and Ambivalent Nature of Robert

Beyond technical analysis, our results suggest that Comment Robert represents a transformative moment in Chinese social media. Unlike traditional AI tools that respond passively, Robert actively participates in ongoing conversations, often without user consent or clear rules of engagement. It disrupts the boundary between user and algorithm, inserting itself into intimate expressions and public debates alike. For many users, Robert is not just a tool but a social actor—sometimes seen as a friend, sometimes as a threat. Its unpredictability makes it entertaining and viral, but also difficult to trust. This blurred identity—neither fully machine nor fully person—raises important questions about digital companionship, human-AI interaction, and the limits of algorithmic empathy.

One particularly concerning finding is the potential for covert verbal harm. While Robert rarely uses overtly aggressive language, our analysis shows that it sometimes mirrors negative emotions or validates harmful ideas through a casual, joking tone. The platform's failure to detect or regulate such behavior points to a larger issue: emotional risk in AI interaction is not just about what is said, but how it is said and how it is perceived.

In conclusion, Robert has sparked a new phase of social experimentation in China's digital landscape. It has become a cultural symbol, a source of entertainment, and a mirror reflecting users' desires, anxieties, and boundaries with technology. But its rise also exposes the risks of deploying emotionally reactive AI into public platforms without adequate safeguards. As chatbots become more embedded in daily life, the challenge ahead is not only technical, but social and ethical: how to build machines that can communicate without causing harm, and how to design algorithms that understand the emotional weight of their words.

### 5.3 Implications for AI Chatbot Design and Platform Governance

The findings of this study carry important implications for the design and management of AI chatbots within social media platforms. As AI becomes more embedded in everyday online interactions, platforms and developers face a growing responsibility to balance innovation with user safety and emotional well-being.

First, transparency and user consent must be prioritized. Robert's unsolicited engagement, while novel and engaging for some, can also be intrusive or unwelcome. Platforms should clearly inform users when they are interacting with AI and provide easy ways to opt out of such interactions. Transparent communication about the chatbot's nature and operational limits will help manage user expectations and reduce feelings of confusion or mistrust.

Second, emotional sensitivity and risk detection need to be significantly enhanced. Current automated tools, including widely used toxicity detection APIs, often fail to capture nuanced emotional harm or covert hostility embedded in seemingly benign or playful language. AI systems should be equipped with more sophisticated context-aware models capable of recognizing indirect aggression, sarcasm, and inappropriate reinforcement of negative emotions. Combining algorithmic detection with human moderation, especially in sensitive or high-risk cases, can better safeguard users' mental health.

Third, platforms should foster community feedback mechanisms. As seen in the formation of the "Robert Victims Alliance," user communities can provide valuable insights into the chatbot's social impact and potential unintended consequences. Creating channels for users to report problematic interactions and participate in iterative chatbot improvements will enhance accountability and responsiveness.

Finally, regulatory frameworks should evolve to address the emerging challenges posed by socially proactive AI social bots. Government agencies and relevant institutions have a critical role to play in setting clear ethical and operational standards for AI deployment in public digital spaces. As AI chatbots become more active participants in digital discourse, regulation must ensure they uphold transparency, user rights, and emotional well-being. Key areas of oversight include informed consent, emotional safety, data privacy, and appropriate responses to harmful interactions. Public oversight and cross-sector collaboration will be essential in balancing innovation with user rights, ensuring that AI technologies contribute positively to digital public life.

In sum, integrating AI social bots into social media requires a multidimensional approach—balancing technical innovation, ethical responsibility, and ongoing user engagement. Only through such comprehensive governance can platforms ensure that AI-enhanced interactions remain safe, respectful, and emotionally supportive for diverse user populations.